\documentstyle[12pt]{article}

\begin{document}

\title{On the Nature of Nonperturbative Effects \\
            in Stabilized 2D Quantum Gravity}

\author{{\bf Oscar Diego}\thanks{e-mail: imtod67@cc.csic.es} \ and \
        {\bf Jos\'e Gonz\'alez} \thanks{e-mail: imtjg64@cc.csic.es} \\
        {\em Instituto de Estructura de la Materia } \\
        {\em Serrano 123, 28006 Madrid} \\
        {\em Spain } }

\date{\mbox{ }}

\maketitle

\thispagestyle{empty}

\begin{abstract}
We remark that the weak coupling regime of the stochastic
stabilization of 2D quantum gravity has a unique perturbative vacuum,
which does not support instanton configurations. By means of Monte
Carlo simulations we show that the nonperturbative vacuum is also
confined in one potential well. Nonperturbative effects can be
assessed in the loop equation. This can be derived from the Ward
identities of the stabilized model and is shown to be modified by
nonperturbative terms.
\end{abstract}

\begin{flushright}
\vspace{-13.5 cm} {IEM-FT-65/92}
\end{flushright}
\baselineskip=21pt
\vfill
\newpage
\setcounter{page}{1}

{\bf 1.  Introduction}

It is well know that discretized 2-D quantum gravity coupled to conformal
matter with $ c < 1 $ can be represented by a zero-dimensional matrix model
\cite{discrete,kazakov,ds}
and that the
expansion in powers of $ 1/N $ of the matrix model, where $ N $ is the
dimension of the hermitian matrix field, defines the topological expansion of
the original discretized 2-D quantum gravity. The partition function of
zero-dimensional matrix model has the form
\begin{equation}
Z = \int d \Phi\, \exp \Bigl\{ - N Tr W ( \Phi ) \Bigr\}
\label{eq:11}
\end{equation}
where $W$ is the potential and $ \Phi $ is an hermitian matrix field. The
continuum limit of the discretized 2-D quantum gravity is achieved by taking
the
limit of the coupling constants of the potential $W$ to some critical set of
coupling constants for which the partition function
(\ref{eq:11}) has non-analytic
behaviour. For pure gravity the
argument of the integral (\ref{eq:11})
is an increasing function for large values of $\Phi$
and the integral does not exist for finite $N$, but in the
large $N$ limit the integral exists up to some value of the coupling constant.
This is the critical value of the coupling constant, and the non-analytic
behaviour developed makes the partition function above it not well-defined.
The matrix model is defined only in the
large $N$ limit and its expansion in powers of $ 1/N $ represents the
topological expansion of the quantum gravity, but the matrix model does not
define uniquely the non perturbative 2-D quantum gravity \cite{dav}.

Stochastic stabilization \cite{gre,amv} provides, on the other hand, a
way of mapping the model (\ref{eq:11}) into one which is defined
for all values of $N$ and the coupling constant, while
reproducing the perturbative expansion in powers of $ 1/N $.
There has been much hope that this stabilization of the original
matrix model could lead to an unambiguous nonperturbative
definition of quantum gravity. If these expectations have not
been met, it has been in part because of the poor understanding
of nonperturbative effects in the stabilized theory. As we
review afterwards, this is defined in one more dimension than
the original matrix model, setting therefore the problem into the
framework of quantum mechanics. In fact, in the zero-dimensional
model (\ref{eq:11}) the source of nonperturbative effects lies
on the transmission of eigenvalues through the walls of the
confining potential well \cite{daz}, so that one
could expect the same kind
of phenomenon in the stabilized theory. The main point of this
article, however, is that the ground state of the
one-dimensional model does not bear tunnelling between different
wells of the potential. We are able to prove this, for the
simplest matrix model, in perturbation theory as well as in the
nonperturbative regime (using a Monte Carlo approach in this
latter case). This leaves open the proper interpretation of the
nonperturbative effects. Their significance can be assessed
otherwise, since, as also shown in the paper, they modify the
form of the loop equation in the stabilized theory.

{\bf 2.  Perturbative approach }

The stochastic stabilization of the matrix model introduces a positive definite
hamiltonian
\begin{equation}
H = \frac{1}{2} Tr \left\{- \frac{1}{N^2}\frac{\partial^2}{\partial\Phi^2} +
\frac{1}{4} \left( \frac{ \partial W }{ \partial \Phi} \right)^{2} -
\frac{1}{2 N} \frac{\partial^2 W}{\partial\Phi^{2}}  \right\}
\label{eq:12}
\end{equation}
This hamiltonian is well defined for all values of $N$ and the coupling
constants.
The zero mode of the stabilized theory is given by
\begin{equation}
\Psi(\Phi) \sim \exp \left\{ - N \frac{W(\Phi)}{2} \right\}
\end{equation}
its norm being the partition function of the original matrix model.
Hence, the original matrix model is well defined if and only if this
zero energy state is a
normalizable state. When this is the case, corresponding obsevables in both
theories coincide
\begin{equation}
\langle Q \rangle_{stab} = \int Q \mid\Psi\mid^2 d\Phi =
\frac{1}{Z} \int Q \exp\Bigl\{- N W (\Phi)\Bigr\} d\Phi =
\langle Q \rangle_{matrix}
\end{equation}
We can use the stabilized theory to calculate the observables of the
original matrix model, with the ground state energy playing the role of an
order parameter which
gives us information about the range of definition of the integral
(\ref{eq:11})
. The non-analytic
behaviour of the original matrix model becomes now non-analytic behaviour
in the ground state energy of the hamiltonian (\ref{eq:12}).
For pure gravity in the large $N$ limit, below the critical point the zero
energy state is a normalizable state and the ground state
energy is zero, while
above the critical point the zero-dimensional matrix model does not exist and
the ground state energy of the stabilized hamiltonian is greater than zero.
The stabilized theory exists anyway for all values of $N$ and the coupling
constant, even if the original matrix model is ill defined.

We spend a few time showing that, despite the strange critical
exponent for the leading contribution \cite{gon}, the
topological expansion
of the ground state energy can be organized in terms of the
scaling variable of the zero-dimensional matrix model. Dealing
with the $1/N$ expansion we will also show our main conclusion in the
weak coupling regime, i.e. that the eigenvalues of $\Phi $ in
the ground state are all confined in the same well of the
potential in (\ref{eq:12}).
Let us consider the cubic matrix model given by
\begin{equation}
W(\Phi) = Tr \Phi^2 - \frac{2 g}{3} Tr \Phi^3 ,
\end{equation}
where $\Phi$ is the $N$
dimensional hermitian matrix. We consider throughout this section the
large $N$ limit of the model, for which (\ref {eq:11}) is still
defined within a certain range at $g>0$.
It is well-known that the one-dimensional matrix model is
mapped into a gas of free fermions \cite{bipz}, which are the eigenvalues
$\{\lambda_{n}\}$ of the $\Phi$ matrix
variable\footnote{In other matrix models,
the hamiltonian of the stabilized theory
has interaction terms, but one can perform a Hartree-Fock approach
to obtain the exact ground state energy in terms of suitable one-particle
states \cite{gon,mir,die}.}.
The Fokker-Planck hamiltonian of the stabilized model (\ref{eq:12})
becomes the sum of $N$ one-particle hamiltonians \cite{gre,amv,amb,amj,amk}
\begin{equation}
h_{n} = - \frac{1}{2} \frac{1}{N^2} \frac{\partial^2}{\partial\lambda_n^2} +
\frac{1}{2} \{ g^2 \lambda_n^4 - 2g\lambda_n^3 + \lambda_n^2 +
2g\lambda_n - 1 \}
\label{eq:21}
\end{equation}
The one-particle hamiltonian (\ref{eq:21}) has a main well
and a local minimun below the critical coupling constant
$g_{c} = \sqrt{1/(6 \sqrt{3})}$,
and only one well
above it (see figure 1).

The ground state energy of the Fokker-Planck hamiltonian is
\begin{equation}
E=\sum_{n=0}^{N-1} e_{n}  \label{tote}
\end{equation}
where $\{e_{n}\}_{0 \leq n \leq N-1}$ are the first $N$ eigenvalues
of the one particle
hamiltonian (\ref{eq:21}).
In the large $N$ limit we can write down for it an integral
representation, taking into account $1/N$ effects. From
the Euler-Maclaurin sum-formula \cite{for} the total energy is
\begin{equation}
E = \sum_{n=0}^{N-1} e_{n} = \int_{E_{0}^{\ast}}^{E_{F}^{\ast}} dE E \rho(E) -
\frac{1}{2} E_{F}^{\ast} + \frac{1}{2} E_{0}^{\ast} + \frac{1}{12} \frac{1}{
\rho(E_{F}^{\ast})} - \frac {1}{12} \frac{1}{\rho(E_{0}^{\ast})} + \cdots
\label{eq:22}
\end{equation}
where $E_{0}^{\ast}$ is the first eigenvalue $ e_{0} $,
$E_{F}^{\ast}$ is the value of the analytic continuation of $e_{n}$ from
$n \leq N-1$ to $n = N$, and $\rho (E)$ is the density of states.

{}From (\ref{eq:21}) we see that $1/N$ play the role of $\hbar$, and
the large $N$ limit is a semiclassical expansion. The density of states
is given by
\begin{equation}
\rho (E) = \frac{d}{dE} N(E)
\end{equation}
The distribution function of states is given by \cite{par}
\begin{equation}
N(E) = \frac {N}{\pi}\int_{a}^{b} d\lambda \sqrt{2(E-V)}
- \frac{1}{24\pi N}  \int_{a}^{b} d \lambda \frac{d^2V}{d\lambda^2}
\frac{1}{\sqrt{2(E-V)}} + O \left( \frac{1}{N^3} \right)
\label{eq:23}
\end{equation}
$a$ and $b$ being  the classical turning points.
$E_{F}^{\ast}$ and $E_{0}^{\ast}$ obey the
quantization conditions
\begin{equation}
N( E_{0}^{\ast} ) = \frac {1}{2}
\end{equation}
and
\begin{equation}
\tilde{N} ( E_{F}^{\ast} ) = N + \frac{1}{2}
\end{equation}
where $\tilde{N}(E)$ is the distribution function $N(E)$ restricted to the
main well of the stabilized potential.
In general, one has to keep track of higher order terms in the
sum formula (\ref{eq:22}) altogether with the perturbative
expansion (\ref{eq:23}), in order to obtain correctly
the $1/N^2$ expansion of the total energy.

{}From (\ref{eq:22}) and (\ref{eq:23}) the total energy is
\begin{equation}
NE = N^{2} \left( E_{0} + E_{2} \frac {1}{N^2} + O \left(
\frac{1}{N^4} \right) \right)
\label{eq:24}
\end{equation}
where
\begin{equation}
E_{0} = E_{F}^{(0)} - \frac{1}{3\pi} \int_{a}^{b} d\lambda
[2(E_{F}^{(0)} - V)]^{\frac{3}{2}}  \label{e0}
\end{equation}
\begin{equation}
E_{2} = \frac{1}{24\pi} \int_{a}^{b} d\lambda \frac{d^2V}{d\lambda^2} \frac{1}
{\sqrt{2(E_{F}^{(0)} - V )}}
- \frac{1}{24} \left[ \frac{1}{\pi} \int_{a}^{b} d\lambda
\frac {1}{\sqrt{2(E_{F}^{(0)} - V )}} \right]^{-1} \label{e2}
\end{equation}
Here $ E_{F}^{(0)} $ is the Fermi energy in the planar limit
and is given by the
condition
\begin{equation}
\frac{1}{\pi} \int_{a}^{b} d\lambda \sqrt{2(E_{F}^{(0)} - V )} = 1
\label{fl}
\end{equation}

We expect the total energy to be zero for $g < g_{c}$ to all
perturbative orders \cite{gre}, which we have actually checked
to second order in $1/N$ expansion. The leading order of the
total energy (\ref{e0}) is zero up to the critical coupling
$g_{c} = \sqrt{1/(6 \sqrt{3})}$, and greater than zero above it.
The critical behaviour of this quantity agrees with the
exponent found in the cuartic model \cite{gon}
\begin{equation}
E_{0}  \sim  (g - g_{c})^{11/4}  \;\;\;\;\;\;\;\;  g > g_{c}
\end{equation}
Regarding the subleading contribution (\ref{e2}) we find the
critical behaviour
\begin{eqnarray}
E_{2}  & = &  0       \;\;\;\;\;\;\;\;\;\;    g < g_{c}  \\
E_{2} & \sim &  (g - g_{c})^{1/4}  \;\;\;\;\;\;\;\;  g > g_{c}
\end{eqnarray}
The appropriate double scaling limit $ g \rightarrow g_{c} $ and
$ N \rightarrow \infty$  is such that $z=N(g_{c}-g)^{\frac{5}{4}}$
remains finite \cite{ds}. We see, in fact, that the topological
expansion of the total energy organizes above the critical point
as a power series in the scaling variable $z$ of the original
matrix model
\begin{equation}
NE = N^{2} ( g - g_{c} )^{\frac{11}{4}} \left( B_{1} + B_{2}
\frac {1}{N^2 ( g - g_{c} )^{\frac{5}{2}}} + \cdots \right)
\end{equation}

The striking result, however, concerns the position of the Fermi
level $E_{F}$. To leading order of the $1/N$ expansion,
$ E_{F}^{(0)} $ is characterized by the condition (\ref{fl}), which places
it precisely at the level of the local minimum of the potential,
all the way up to $g_{c}$ \cite{mar}. The first order correction
$ E_{F}^{(1)} $ can be computed from the quantization condition
\begin{equation}
\frac{N}{\pi} \int_{a}^{b} d\lambda \sqrt{2(e_{n} - V )} +
O\left( \frac{1}{N} \right)  = n + \frac{1}{2}
\label{qu}
\end{equation}
bearing in mind that we have to fill the first $N$ levels from
$e_{0}$ to $e_{N-1}$ as in (\ref{tote}). We find then
\begin{equation}
E_{F}^{(1)} = - \frac{1}{2} \left[ \frac{1}{\pi} \int_{a}^{b} d\lambda
\frac{1}{\sqrt{2(E_{F}^{(0)} - V )}} \right ]^{-1}
\end{equation}
which is a negative quantity. This means that, at least in the
weak coupling regime, the ground state appears to be confined to
the central well of the potential, and that there is no issue
about tunnelling to the region around the local minimum.
One may think that this only points at the
unfeasibility of discussing nonperturbative effects in the very
framework of the $1/N$ expansion. Actually, this approximation
implies taking the limit $N \rightarrow \infty$ from the start,
so that it would be conceivable that a more sophisticated way of
tuning the double scaling could lead to a different physical picture.
In the last section we will return to the question of the
localization of the ground state of the model at finite $N$.

{\bf 3.  Observables and the loop equation}

Below the critical point, the observables of the stabilized theory
and the matrix model have to be the same, and may be calculated
as follows \cite{amb,die}.

We add to the potential $V$ an auxiliary
term $\beta \lambda^{n}$ and, using
the Hellmann-Feynman theorem, write
the following formula for observables of
the stabilized theory
\begin{equation}
\frac{1}{N} \langle tr \Phi^{n} \rangle =
\left ( \frac{\partial E (\beta) }
{\partial \beta } \right)_{\beta = 0}.
\label{eq:31}
\end{equation}

Taking into acount (\ref{eq:24}), (\ref{eq:31}) and results from appendix A,
we perform the calculation of the observables of the theory up
to second order in $\frac{1}{N}$. To first order all observables
are given by
\begin{equation}
\langle tr \Phi^{n} \rangle = N^{2} \int_{a}^{b} d
\lambda \lambda^{n} \rho(\lambda)
\end{equation}
and
\begin{equation}
\rho (\lambda) = \frac{1}{\pi} \sqrt{2(E_{F}^{0} - V )}
\end{equation}
is the fermionic density and is equal to the eigenvalue density of the
original matrix model in the planar limit for $g < g_{c}$ \cite{mir}.
Now we perform the limit $ g \longrightarrow g_{c}$, which represents
the continuum limit of the discretized 2-D quantum gravity,
and calculate numerically the observables. After computing
the most singular part of the observables turns out to be given by
\begin{equation}
\langle tr \Phi^{n} \rangle = N^{2} (g_{c} - g)^{\frac{3}{2}}
\left ( A_{1}^{(n)} + A_{2}^{(n)} \frac{1}{N^{2} ( g_{c} - g )
^{\frac{5}{2}}} + \cdots \right)
\label{eq:32}
\end{equation}
for $g < g_{c}$, and
\begin{equation}
\langle tr \Phi^{n} \rangle = N^{2} (g - g_{c} )^{\frac{3}{2}}
\left ( C_{1}^{(n)} + C_{2}^{(n)} \frac{1}{N^{2} ( g - g_{c} )
^{\frac{5}{2}}} + \cdots \right)
\label{eq:33}
\end{equation}
for $g > g_{c}$.

In the double scaling limit (\ref{eq:32}) defines the puncture
operator of the 2D quantum gravity. As long as the critical exponents
in expresions (\ref{eq:32}) and (\ref{eq:33}) are the same, we can
define an analogous double scaling limit above the critical point given
by $ z = N ( g - g_{c})^{\frac{5}{4}} $. This limit does not
exist in the zero-dimensional matrix model, hence, it defines new effects
beyond the original formulation of matrix model. In reference \cite{amj}
this limit is used to show that the stabilized model does not
satisfy the nonperturbative KdV flow equations.

The loops equations arise from the stabilized model as follows:
We add to the potential $W$ a perturbation
\begin{equation}
W = Tr \Phi^2 - \frac{2g}{3} Tr\Phi^3 - \frac{2 \beta_n}{n} Tr \Phi^n .
\end{equation}
The potential of the stabilized theory becomes now
\begin{eqnarray}
V ( \beta_n) & = & \frac{1}{2} \{ Tr\Phi^{2} + g^{2}Tr\Phi^{4} - 2gTr\Phi^{3} +
2gTr\Phi - N \nonumber  \\
& & \mbox{} + \beta_{n}^{2} Tr\Phi^{2n-2} - 2\beta_{n}Tr\Phi^{n} + 2g\beta_{n}
Tr\Phi^{n+1} \nonumber \\
& & \mbox{} + \frac {\beta_n}{N} \sum_{k=0}^{n-2} Tr\Phi^{k} Tr\Phi^{n-2-k} \}
\end{eqnarray}
now from the Hellmann-Feynmam theorem
\begin{equation}
\left( \frac{\partial E }{\partial \beta_n }
\right)_{\beta_n = 0 } = - \langle
Tr\Phi^n\rangle + g \langle Tr\Phi^{n+3} \rangle + \frac{1}{2N}
\sum_{k=0}^{n-2} \langle Tr \Phi^k Tr \Phi^{n-2-k} \rangle.
\end{equation}

The loop of lenght $L$ is given by
\begin{equation}
W(L) = \frac{1}{N} Tr e ^{L\Phi}
\end{equation}
and is not difficult to prove that
\begin{equation}
\dot {V} \left(\frac{\partial}{\partial L } \right) \langle W (L) \rangle -
\int_{0}^{L} dJ \langle W(J) W( L - J ) \rangle =
- \frac{2}{N} \sum_{n=0}^{\infty} \frac{L^n}{n!}
\left (\frac{\partial E}
{\partial \beta_{n+1}} \right ) _{\beta_{n+1} = 0}
\label{eq:ab1}
\end{equation}
then, if the energy of the ground state is zero for all values of $\beta_{n}$
the Hellmann-Feynmam theorem gives the set of Ward identities
\begin{equation}
\left ( \frac{\partial E}{\partial \beta_{n}} \right )_{\beta_{n} = 0}
= 0  \  \  n = 0,..., \infty
\end{equation}
which are equivalent to the first loop equation
\begin{equation}
\dot {V} \left(\frac{\partial}{\partial L } \right) \langle W (L) \rangle =
\int_{0}^{L} dJ \langle W(J) W( L - J ) \rangle.
\end{equation}
We expect that the other loop equation can be found when we add to the
potential $W$ perturbation like :
\begin{equation}
W = Tr \Phi^2 - \frac{2g}{3} Tr\Phi^3 - \frac{2 \beta_{n_1, n_2,... }}{n_1
n_2...} \prod_{i} Tr \Phi^{n_i}.
\end{equation}

The stabilized hamiltonian (\ref{eq:12})
is the supersymmetric hamiltonian of
reference \cite{mar} restricted to the bosonic sector.
The Ward identities of the supersymetric matrix model becomes
the Swchinger Dyson equation of the zero-dimensional matrix model.

In the case of pure gravity the supersymmetry is broken non perturbatively,
then the Ward identities or the loop equation are modified by non
perturbative corrections.

{\bf 4.  Nonperturbative approach  }

We come back to the question of the localization of the ground state
in the stabilized theory, working near the critical coupling at
finite values of $N$. In order to get
information about the eigenvalue
distribution of the ground state, we apply the Monte Carlo
method in the computation of the path integral for the quantum system.
In reference \cite{amv} a Monte Carlo
method  was also applied to simulate observables of the  matrix
variable $ \Phi $. We consider here an alternative approach  which
optimizes the Monte Carlo calculation reducing the number of
variables to the eigenvalues of $ \Phi $.

The basic object to look for is the probability amplitude
between two states $\Psi_{i} $ and $\Psi_{f} $ at times $0$ and
$T$, respectively. This admits the path integral
representation \cite{path}
\begin{eqnarray}
\langle \Psi_{f}(T) |  \Psi_{i}(0) \rangle
 & = & \int\prod_{i=1}^{N}D\lambda_{i}(t) \Delta(\lambda(0))
\Delta(\lambda(T))  \overline{\Psi}_{f} (\lambda (T) )
   \Psi_{i} (\lambda (0) ) \nonumber \\
        &   & \exp{\left \{-N\int_{0}^{T}dt
\sum_{n=1}^{N}(\frac{1}{2}\dot{\lambda_{n}}(t)+
\frac{1}{2}V_{FP}(\lambda_{n}(t)))\right\}}
\label{eq:41}
\end{eqnarray}
where $\Delta(\lambda)$ is the Van der
Monde determinant
 and $V_{FP}$  the stabilized potential
appearing in (\ref{eq:21}). If both $\Psi_{i}$ and $\Psi_{f}$
have nonvanishing projection over the ground state, this is the
state which dominates at large $T$. The
amplitude behaves, in terms of the ground state energy $E$,
\begin{equation}
\langle \Psi_{f}(T) |  \Psi_{i}(0) \rangle
  \sim  \mbox{\Large $e^{- E T}$}
\end{equation}

We have performed, in practice, a discretization of the time $T$,
such that
$ t_{j}  = \epsilon \ j $,
where $ j = 0, \cdots, N_{T} $ and $ t_{N_{T}} = \epsilon N_{T} = T$.
The measure of integration in (\ref{eq:41}) becomes
\begin{eqnarray}
\mbox{\Large $e^{-S}$}
  & = & \exp \left\{- N \sum_{i=1}^{N} \sum_{j=0}^{N_{T}-1}
\left ( \frac{1}{2}
\frac{(\lambda_{i}(t_{j+1}) - \lambda_{i}(t_{j}) )^{2}}{\epsilon} + \epsilon
V_{FP}
(\lambda_{i}(t_{j})) \right )  \right.  \nonumber \\
 &  & +  \sum_{i < j} \log{(
\lambda_{i}(t_{0}) - \lambda_{j}(t_{0}) )}  \nonumber \\
 &  &  \left. +  \sum_{i < j} \log{(
\lambda_{i}(t_{N_{T}}) - \lambda_{j}(t_{N_{T}}) )}  \right\} \label{expo}
\end{eqnarray}
This can be simulated by the Monte Carlo method, as if we were
dealing with a two-dimensional statistical system of size
 $ N \times N_{T} $ . The variables are the
eigenvalues $\lambda_{i} ( t_{j} ) $ where $ i = 1, \ldots, N $,
$ j = 0, \ldots, N_{T} $.
The kinetic term defines a nearest-neighbor interaction along the time
axis and the determinants in (\ref{eq:41}) ---the only vestige
of Fermi statistics in the path integral--- define a long-range interaction
at the
boundaries $t = 0$ and $t = T$. There are no interactions between bulk
variables along the $N$ axis.

The crucial point in the simulation is to reach a large enough
value of the time $T$, in order to measure with confidence
the ground state properties. We have taken steps with $\epsilon
= 0.1$, and found that above $T = 8$ observables measured with
(\ref{expo}) do not show significant contribution from the first
excited states. We have imposed boundary conditions in the form
of constant wave functions at both ends of the time interval
$\Psi_{i}(\lambda) = \Psi_{f}(\lambda)
= const. $ , which have
certainly nonvanishing overlap with the ground state.

We have implemented the Monte Carlo simulation in a VAX 9000 machine
with Metropolis algorithm. We have made a random
choice of the point $t_{i}$ each time, updating then the $N$
variables at that site.
The number of iterations between measures has been $ 2000 $ MC sweeps
and we have left a thermalization period equivalent to
$ 5 \times 10^{4} $ MC sweeps of all the variables in the lattice.
We have simulated systems with $N = 5$ and $N = 10$, starting
with different initial conditions for the set of eigenvalues.
In some of the simulations
part of the eigenvalues were initially confined
in the well around the local minimum and in the others
all of them were in the central well. Irrespective of these
different choices, after the thermalization period we ended up
with a distribution of eigenvalues equal to zero in the well
around the local minimum. This happened even for values of $g$ very
close to $g_{c}$. Figure 2 shows a typical distribution for $N =
5$ and $g - g_{c} = 10^{-4}$ after $ 8 \times 10^{6} $ MC sweeps
of all the lattice. This supports strong evidence that at finite
$N$ the Fermi level is placed below the level of the local
minimum of the potential, all the way up to $g_{c}$.

{\bf 5.  Conclusions  }

We have shown that in the framework of the $1/N$ approximation
there is only one perturbative
vacuum. The Fermi level in the weak coupling regime is below the
local minimum of the
potential and,
hence, instanton configurations which start or end
at the local minimun do not exist.
In the nonperturbative approach we have also made plausible that the
ground state does not bear tunnelling between different wells of
the potential.
The explanation of this picture should be the following.
The stochastic stabilization
may be viewed as the usual stochastic
quantization with asymptotic final and initial states fixed and
given by well defined configurations of the original
model \cite{grep}. Therefore, in the stabilized matrix model
all configurations have to start and end
at the main well of the stabilized potential. The Fermi level
has to be
placed below the local minimun and an instanton which starts or ends at
the local minimun cannot exist.

One possible way of understanding the nonperturbative effects
may be envisaged as follows.
There are two kind of static solutions of the classical
equation in the large $N$ limit: (a)
all the fermions are restricted to the main
well of the potential, and (b) a fermion is placed at the local minimun.
These
two solutions are degenerate at first order in $1/N$.
Hence we expect that the time dependent solution of the
classical equation of motion connects the main well and the local minimun.
The perturbative effects lift the degeneracy, and the solution with a
fermion in the local minimun becomes an excited state. The nonperturbative
effects have to arise from trajectories of one fermion
which start and end at the main well.
These closed trajectories
are given by a sucession of instantons
and anti-instantons \cite{zin}. Hence, the instantonic action of one fermion
is
\begin{equation}
S_{inst} = 2 \int_{a}^{b} \sqrt{2(V-E_{F})}
\label{eq:c1}
\end{equation}
which agrees with the instantonic action calculated in \cite{mir}.

In this paper we have considered the double scaling limit in the order: first
take
the large $N$ limit and then $ g \rightarrow g_{c} $. This is the only double
scaling limit defined in the zero-dimensional matrix model.
But in the stabilized model it is
possible to perform the double scaling limit from finite values of $N$ and $g -
g_{c}$. We have performed a preliminary numerical calculation.
In order to understand how an alternative double scaling limit
may be  achieved from
finite values of $N$, a more detailed investigation of the
phases in the space of parameters $(N, g)$ should be carried out.

\newpage

{\bf    Appendix A}

In this appendix we show how formulas like
\begin{equation}
\frac{\partial}{\partial\beta} \int_{a}^{b} d\lambda
\frac{1}{\sqrt{2(E_{F}^{(0)} -
V)}}
\end{equation}
which have end points singular integrand, can be calculated.

We define two fixed arbitrary points $\Lambda_1$ and $\Lambda_2$ such that
\begin{eqnarray}
\int_a^b d\lambda \frac{1}{\sqrt
{2(E_{F}^{(0)} - V)}} & = & \int_a^{\Lambda_2} d\lambda
\frac{1}{\sqrt{2(E_{F}^{(0)} - V)}}
+ \int_{\Lambda_2}^{\Lambda_1} d\lambda
\frac{1}{\sqrt{2(E_{F}^{(0)} - V)}} \nonumber  \\
& & \mbox{} + \int_{\Lambda_1}^{b}
d\lambda \frac{1}{\sqrt{2(E_{F}^{(0)} - V)}}
\end{eqnarray}

For $g\neq g_{c}$, $\frac{dV}{d\lambda}$ has
only one
zero in the interval $[a, b]$, then we choose $\Lambda_1$ and $\Lambda_2$
such that the zero of $\frac{dV}{d\lambda}$ is placed between
$\Lambda_1$ and $\Lambda_2$, see $(fig. 1,a)$.

The second integral can be derived directly
\begin{equation}
\frac{\partial}{\partial\beta}\int_{\Lambda_2}^{\Lambda_1} d\lambda
\frac{1}{\sqrt{2(E_{F}^{(0)} - V)}} =
\int_{\Lambda_2}^{\Lambda_1} d\lambda
\frac{\partial}{\partial\beta}\left\{\frac{1}
{\sqrt{2(E_{F}^{(0)} - V)}}\right\}
\end{equation}
and the result is well defined for all values of the coupling constant.

The other integrals are posibly singular and we perform them as follows
\begin{eqnarray}
\int_{\Lambda_1}^{b} d\lambda \frac{1}{\sqrt{2(E_{F}^{(0)} - V)}} & = &
\int_{\Lambda_1}^{b} d\lambda \frac{1}
{\sqrt{2(E_{F}^{(0)} - V)}} \frac{\dot{V}}{\dot{V}} \nonumber\\
 & = & - \int_{\Lambda_1}^{b} d\lambda \frac{1}{\dot{V}}
\frac{d}{d\lambda}(\sqrt{2(E_{F}^{(0)} - V)})
\end{eqnarray}
where the dots are $\lambda$ derivatives, and integrations by parts gives
\begin{equation}
\int_{\Lambda_1}^{b} d\lambda \frac{1}{\sqrt{2(E_{F}^{(0)} - V)}} =
\ Analytic \ terms \ + \int_{\Lambda_1}^{b} d\lambda
\frac{d}{d\lambda} \left (
\frac{1}{\dot{V}} \right ) \sqrt{2(E_{F}^{(0)} - V)}
\end{equation}
the result is well defined for $g\neq g_{c}$ and can be derived directly, but
if $g = g_{c}$ then $\dot{V} (b) = 0$ and the integral is posibly singular.
This is the origin of the critical point $g_c$, which is defined by the
condition $\dot{V} (b) = 0$, see $( fig. 1,c )$.

To all orders in $1/N$, the energy and observables are some
combinations of integrals which have the form
\begin{equation}
\frac{d^{j}}{d\beta^{j}}\frac{d^{k}}{dE^{k}}
\left( \int_{a}^{b} d\lambda \frac{P(\lambda)}
{\sqrt{2(E - V(\beta))}} \right)_{E=E_{F}^{(0)},\beta=0}
\end{equation}
This integrals have the same critical point $g_{c}$. The critical
point $g_{c}$ remains constant to all orders in $1/N$ expansion.

\vfill

\pagebreak

\pagebreak

\newpage

\centerline{\bf Figure Captions}

\begin{itemize}

\item[Figure 1.] Plot of the stabilized potential a) below the critical point,
b) above the critical point and c) at the critical point

\item[Figure 2.] Plot of the normalized fermionic density:
continuun line.
For $N = 5$, $g - g_{c} =
10^{-4}$, $100$ time intervals . We have performed $4200$ measures.
The planar fermionic density is given by
the dashed line. The vertical dashed
line is placed at the absolute
minimun of the stabilized potential. The local maximun and
the cut of the support of $\rho$ are very near of the local
minimun.

\end{itemize}

\end{document}